\documentstyle[aps,amsfonts,amssymb,floats,epsf]{revtex}
\begin{document}
\draft
\twocolumn[\hsize\textwidth\columnwidth\hsize\csname@twocolumnfalse\endcsname

\title{Network Model for a 2D Disordered Electron System with
  Spin-Orbit Scattering}
\author{Rainer Merkt, Martin Janssen, and Bodo
  Huckestein}
\address{Institut f\"ur Theoretische Physik, Universit\"at zu
  K\"oln, Z\"ulpicher Strasse 77, 50937 K\"oln, Germany }
\date{\today}
\maketitle

\begin{abstract}
  We introduce a network model to describe two-dimensional disordered
  electron systems with spin-orbit scattering. The network model is
  defined by a discrete unitary time evolution operator. We establish
  by numerical transfer matrix calculations that the model exhibits a
  localization-delocalization transition. We determine the
  corresponding phase diagram in the parameter space of disorder
  scattering strength and spin-orbit scattering strength. Near the
  critical point we determine by statistical analysis a one-parameter
  scaling function and the critical exponent of the localization
  length to be $\nu=2.51\pm 0.18$. Based on a conformal mapping we
  also calculate the scaling exponent of the typical local density of
  states $\alpha_0=2.174 \pm 0.003$.
\end{abstract}

\vskip2pc]

\section{Introduction}
Recently, localization-delocalization (LD) transitions in 2D
disordered electron systems in the absence of a magnetic field were
observed by several groups
\cite{KKFPI94,KMBFPI95,KSSMF96,PBPB97,SiKrSa97b,PoFoWa97,SHPLRR97}.
These results are in contrast with the scaling theory for
non-interacting electrons \cite{AALR79}, which predicts that all
states are localized in two dimensions and in the absence of
spin-orbit interaction (SOI). Now, a new discussion has started on
this topic with the emphasis on the effects of electron-electron
interaction and spin-orbit interaction
\cite{DAMC97,ChYiAb97,Pudal97,CaDiLe98,Lyanda98}.

It is known that both types of interactions could be responsible for
the existence of a LD transition. In the case of SOI, general
arguments \cite{berg82} and perturbation theoretical calculations in
the weakly disordered regime \cite{HiLaNa80,MaeFuk81} yield a positive
correction to the conductance. This quantum interference effect
requiring time reversal invariance is known as weak anti-localization.
In the present work we focus on the detailed examination of a 2D
non-interacting electron system with SOI. For these purposes we
formulate a scattering
theoretical network model for such a system.

\begin{figure}[!btp]
  \begin{center}
    \leavevmode
    \epsfxsize=7cm
    \epsffile{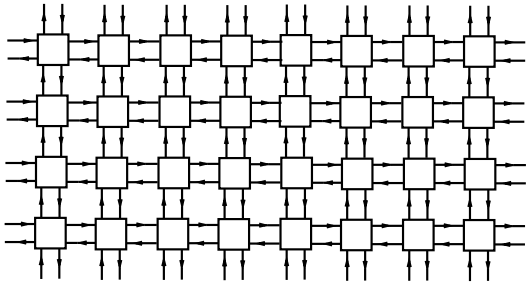}
  \end{center}
  \caption{Topology of a general network model. Squares are scatterers 
    and lines are bonds.}
  \label{PIC-simple-NWM}
\end{figure}

In a recent paper \cite{FrJaMe97} is was shown that scattering
theoretical network models (NWMs) are well suited to describe
mesoscopic disordered electron system. In general such a NWM can
represent any system of coherent waves propagating through disordered
media. It consists of a network of unitary scatterers connected by
bonds.  The arrangement of scatterers and bonds defines the topology
of the NWM, which can be described by a connectivity matrix.  In our
work we have chosen a simple case, where the scatterers are located
on the sites of a quadratic grid, so each of them has four nearest
neighbors. Each bond consists of $2n$ links, $n$ for each direction,
where $n=1$ for waves without and $n>1$ for waves with internal
degrees of freedom (cf. Fig. \ref{PIC-simple-NWM}). In the case of
electron waves a complex number is attached to each link representing
the probability amplitude at this position. The set of all amplitudes
defines the quantum mechanical state $\Psi(t)$ at time $t$. One step
of time evolution is then given by a unitary operator $\cal{U}$,
\begin{equation}
  \Psi(t+1) = {\cal U}\Psi(t).
\end{equation}
This time evolution operator is determined by all the scatterers in
the NWM. Each scatterer maps $4n$ incoming channels to $4n$ outgoing
channels conserving the current and is therefore represented by a
unitary $4n\times4n$-matrix. The disorder is in general simulated in
two ways: first by multiplying the amplitude on each link with a
complex random phase factor ${\text e}^{{\text i}\phi}$ with $\phi$
randomly chosen from $[0,2\pi[$ simulating the random distances
between the scatterers and secondly by taking random values for the
parameters that parameterize the matrix representation of the
scatterers simulating the random strengths of the scatterers. Of
course, both random choices have to be compatible with the symmetry
properties of the system. We distinguish 2D electron systems with time
reversal symmetry (O2NC) and without time reversal symmetry (U2NC),
both without spin degrees of freedom, and systems with time reversal
symmetry and spin degrees of freedom (S2NC). The "2" refers to the
space dimension, the "O", "U" and "S" mean "orthogonal", "unitary" and
"symplectic", which refers to the corresponding universality classes
of random matrix theory and the letters "NC" indicate that all these
systems are "non-chiral", which means that no orientation is preferred
as would be in presence of a strong magnetic field.

The two former models have been examined extensively in
\cite{frech97}. In this work a reflection, a transmission and a
deflection coefficient were introduced, which parameterize the
scattering matrices.  Furthermore an elastic mean free path was
defined in terms of these coefficients. It was concluded that all
states are localized in the O2NC-/U2NC-NWM.

In the present work we investigate the S2NC-NWM. We find a
parameterization for the matrix representation of the spin scatterers
and introduce a spin scattering strength. The calculation of the
localization length by the transfer matrix method allows us to detect
the LD transition and to determine the scaling function. In order to
quantify the scaling exponent $\nu$ of the correlation length we use a
fit procedure \cite{hucke90}. We fit the scaling function and the
critical exponent in two steps respecting the correlations of the
data. Additionally, we apply a $\chi^2$-test to estimate the
confidence of the fits. Determining the critical value of the
localization length we find the scaling exponent $\alpha_0$ of the
typical local density of states using a conformal mapping
\cite{jans94}.

This paper is organized as follows: In Sec. \ref{NWM} we introduce the
network model by explicitly constructing the scattering matrices.
Sec. \ref{FSS} contains the transformation to the transfer matrices
and summarizes general aspects of LD transitions. A detailed
description of the methods of data evaluation forms the content of
Sec. \ref{MOE}. The discussion of the results is presented in Sec.
\ref{RAD} followed by a short summary in Sec. \ref{SUM}.

\section{Network Model}
\label{NWM}
\subsection{Topology}

\begin{figure}
  \begin{center}
    \leavevmode
    \epsfxsize=7cm
    \epsffile{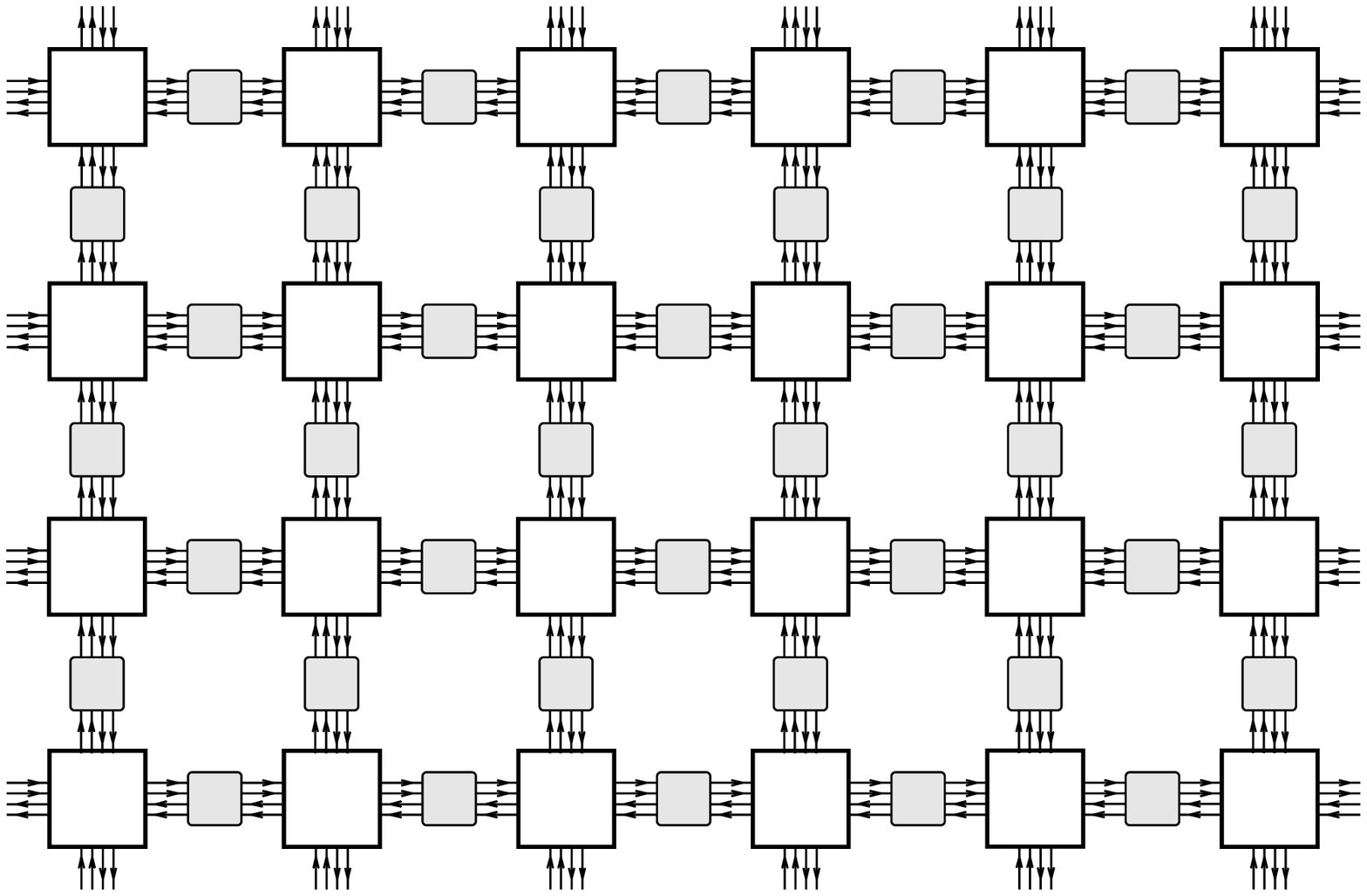}
  \end{center}
  \caption{Topology of the network. The potential scatterers (white) change
    the direction, the spin scatterers (grey) the spin of the electrons.}
  \label{PIC-net-spo-spi}
\end{figure}

There are two different types of scatterers in the S2NC-NWM: {\it
  potential scatterers} (PSs) changing only the electron's direction
and {\it spin scatterers} (SSs) changing only the electron's spin. The
network consists of a regular 2D quadratic grid of potential
scatterers each of them connected to the four next neighbors by
bonds. On each bond a spin scatterer is placed leaving the electron's
direction unchanged (cf. Fig. \ref{PIC-net-spo-spi}).

\subsection{Potential scatterers}
\begin{figure}
  \begin{center}
    \leavevmode
    \epsfxsize=4.5cm
    \epsffile{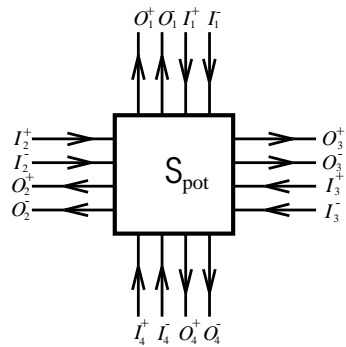}
  \end{center}
  \caption{Potential scatterer: The eight incoming channels
    $I_i^\sigma$ are mapped to the eight outgoing channels
    $O_i^\sigma$.}
  \label{PIC-spo}
\end{figure}

There are four channels or {\em links} within each bond, two of them
for incoming and two of them for outgoing states with spin up and spin
down, respectively. The electronic state is represented by complex
numbers (amplitudes) on each link. Consequently, each PS maps eight
incoming channels $I^\sigma_i$ to eight outgoing channels $O^\sigma_i$
($\sigma\in\{+,-\}$, $i\in\{1,2,3,4\}$) and thus can be represented by
a $8\times8$-matrix ${\sf S}_{\text{pot}}$. With the definition of the
geometrical arrangement of the channels shown in Fig. \ref{PIC-spo}
this mapping is defined as follows:
\begin{equation}
  \bbox{O} = {\sf S}_{\text{pot}} \bbox{I}
  \qquad \text{with} \quad 
  \bbox{I} = 
  \footnotesize
  \left(\begin{array}{c}
    I^+_{1} \\ I^-_{1} \\ \vdots \\  I^+_{4} \\ I^-_{4}
  \end{array}\right)
  \,,
  \normalsize
  \bbox{O} = 
  \footnotesize
  \left ( \begin{array}{c}
    O^+_{1} \\ O^-_{1} \\ \vdots \\ O^+_{4} \\ O^-_{4}
  \end{array} \right ).
  \label{Streu-Allg}
\end{equation}

Due to current conservation,
\begin{equation}
  \sum_{i,\sigma} \left | I_i^\sigma \right |^2 = \sum_{i,\sigma}
  \left | O_i^\sigma \right |^2,
\end{equation}
each scattering matrix has to be unitary,
\begin{equation}
  {\sf S}_{\text{pot}}\cdot{\sf S}_{\text{pot}}^\dagger = \openone_{8},
\end{equation}
where $\openone_8$ denotes the $8\times8$ identity matrix. Additionally,
the scatterers are time reversal invariant. Both properties yield the
matrix to be symmetric,
\begin{equation}
  {\sf S}_{\text{pot}} = {\sf S}_{\text{pot}}{^{\text{T}}},
\end{equation}
where T denotes the transpose.

\begin{figure}
  \begin{center}
    \leavevmode
    \epsfxsize=4cm
    \epsffile{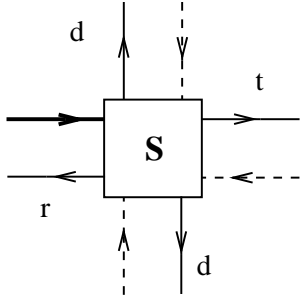}
  \end{center}
  \caption{Definition of the reflection coefficient $r$, the
    transmission coefficient $t$ and the deflection coefficient $d$.}
  \label{PIC-rtd_define}
\end{figure}

For convenience we choose the potential scatterers to be isotropic,
i.e. they are invariant under rotations by multiple angles of $\pi/2$.
With these restrictions each scattering matrix ${\sf S}_{\text{pot}}$
can be parameterized in the following way \cite{frech97}:
\begin{equation}
  {\sf S}_{\text{pot}} = \Phi\tilde{\sf S}_{\text{pot}}\Phi
\end{equation}
with
\begin{equation}
  \tilde{\sf S}_{\text{pot}} = 
  \left ( \begin{array}{cccc}
    r{\text e}^{{\text i}\phi_r} & d & d & t{\text e}^{{\text i}\phi_t} \\
    d & r{\text e}^{{\text i}\phi_r} & t{\text e}^{{\text i}\phi_t} & d \\
    d & t{\text e}^{{\text i}\phi_t} & r{\text e}^{{\text i}\phi_r} & d \\
    t{\text e}^{{\text i}\phi_t} & d & d & r{\text e}^{{\text i}\phi_r} \\
  \end{array} \right )\otimes\openone_2
\end{equation}
and
\begin{equation}
  \Phi =
  \left ( \begin{array}{cccc}
    {\text e}^{{\text i}\phi_{1}} & 0 & 0 & 0 \\
    0 & {\text e}^{{\text i}\phi_{2}} & 0 & 0 \\
    0 & 0 & {\text e}^{{\text i}\phi_{3}} & 0 \\
    0 & 0 & 0 & {\text e}^{{\text i}\phi_{4}} \\
  \end{array} \right )
  \otimes\openone_2.
  \label{S-Pot}
\end{equation}
Here $\openone_2$ denotes the $2\times2$ identity matrix and
$\otimes$ is the tensor product. The real parameters $r,t,d$ denote
the {\em reflection}, {\em transmission} and {\em deflection} (right
and left scattering) {\em coefficient}, respectively (cf. Fig.
\ref{PIC-rtd_define}). If we choose $r$ and $t$ as independent
parameters for ${\sf S}_{\text{pot}}$, the real phases $\phi_r$,
$\phi_t$ and the deflection coefficient $d$ are related to them due to
unitarity and time reversal symmetry,
\begin{mathletters}
  \begin{eqnarray}
    \left | r \right |^{2} + 2\left | d \right |^{2} + \left | t
    \right |^{2} & = & 1,
    \label{rtd-Beding-a}\\ 
    \left | r \right |\cdot\left | d \right |\cos\phi_{r} & = &
    -\left | t \right |\cdot\left | d \right |\cos\phi_{t},\\ 
    \left | r \right |\cdot\left | t \right |\cos(\phi_{r}-\phi_{t})
    & = & \left | d \right |^{2}.
  \end{eqnarray}
  \label{rtd-Beding}
\end{mathletters}
Furthermore, two restrictions follow from these equations,
\begin{mathletters}
  \begin{eqnarray}
    r^{2} +t^{2} \;&\leq\; 1,\\ 
    r\phantom{^{2}} + t \phantom{^{2}} \;&\geq \;1.
  \end{eqnarray}
  \label{r-t-restriction}
\end{mathletters}

The four real phases $\phi_1,\dots,\phi_4$ which are randomly chosen
from the interval $[0,2\pi[$ model the spatial disorder. They can be
interpreted as phase factors ${\text e}^{{\text i}\phi_i}$ for freely
propagating electron waves. Consequently, there are six independent
parameters. But only $r$ and $t$ govern the macroscopic properties of
the system. For convenience we choose them to be equal for all PSs in
the network, whereas the phases are randomly taken for each scatterer.

\subsection{Spin scatterers}
\label{spin_scat}
\begin{figure}
  \begin{center}
    \leavevmode
    \epsfxsize=4.5cm
    \epsffile{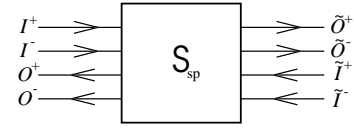}
  \end{center}
  \caption{Spin scatterer: The four incoming channels $I^\sigma$,
    $\tilde I^\sigma$ are mapped to the eight outgoing channels
    $O^\sigma$, $\tilde O^\sigma$.}
  \label{PIC-spi}
\end{figure}

The spin scatterers located between the potential scatterers have two
incoming and two outgoing channels on the left and on the right,
respectively, as is shown in Fig. \ref{PIC-spi}. They can be
represented by $4\times4$-matrices ${\sf S}_{\text{sp}}$:
\begin{equation}
  \bbox{O} = {\sf S}_{\text{sp}} \bbox{I}
  \qquad \text{with} \quad 
  \bbox{I} = 
  \footnotesize
  \left ( \begin{array}{c}
    I^+ \\ I^- \\ \tilde I^+ \\ \tilde I^-
  \end{array} \right )
  \,,
  \normalsize
  \bbox{O} = 
  \footnotesize
  \left ( \begin{array}{c}
    O^+ \\ O^- \\ \tilde O^+ \\ \tilde O^-
  \end{array} \right ).
\end{equation}

Unitarity and time reversal symmetry result in
\begin{equation}
  {\sf S}_{\text{sp}} = {\sf D}^{\text{T}}{\sf
    S}_{\text{sp}}^{\text{T}}{\sf D}
  \label{SP-UNI}
\end{equation}
and
\begin{equation}
  {\sf S}_{\text{sp}} = {\sf D}^{\text{T}}K{\sf
    S}_{\text{sp}}^{-1}K{\sf D} = {\sf D}^{\text{T}}\bigl ({\sf S}_
  {\text{sp}}^{-1}\bigr)^{\ast}{\sf D},
  \label{SP-TR}
\end{equation}
respectively. Here the asterisk denotes complex conjugation and
$K{\sf D}$ is the time-reversal operator with complex conjugation
operator $K$ and
\begin{equation}
  {\sf D} = 
  \left ( \begin{array}{cc}
    -\tau_2 & 0 \\ 0 & -\tau_2
  \end{array} \right )
  =
  \left ( \begin{array}{cc}
    {\sf J} & 0 \\ 0 & {\sf J}
  \end{array} \right )\;,
  \label{D-DEF}
\end{equation}
where $\tau_2$ is one of the basis quaternions
$(\tau_0,\bbox{\tau})=(\tau_0,\tau_1,\tau_2,\tau_3)$ given by
\begin{equation}
  \tau_0 = \openone_2 \quad \text{and} \quad \bbox{\tau} = -{\text
    i}\bbox{\sigma}
\end{equation}
with Pauli matrices $\bbox{\sigma}=(\sigma_x,\sigma_y,\sigma_z)$. The
matrix
\begin{equation}
  {\sf J} = \left ( \begin{array}{cc} 0 & 1\\ -1 & 0 \end{array}
  \right )
\end{equation}
is the symplectic unit matrix. The symmetries (\ref{SP-UNI}) and
(\ref{SP-TR}) suggest the following parameterization of the spin
scattering matrix,
\begin{equation}
  {\sf S}_{\text{sp}} =  
  \left ( \begin{array}{cc}
    0 & {\text e}^{{\text i}{\varphi}}q \\ {\text e}^{{\text
        i}{\varphi}}\bar q & 0
  \end{array} \right )
\end{equation}
with the {\em quaternion real} matrix
\begin{equation}
  q =
  \sum_{k=0}^3 q_k\tau_k=
  \left ( \begin{array}{cc}
    q_{0}-{\text i} q_{3} & -q_{2}-{\text i} q_{1} \\ q_{2}-{\text i}
    q_{1} & q_{0}+{\text i}
    q_{3}
  \end{array} \right )
  \in SU(2),
  \label{Q-Matrix}
\end{equation}
where the real coefficients $q_i$ are restricted by
$\sum_{k=0}^{3}q_{k}^2=1$ due to unitarity and $\bar q$ denotes the
quaternion conjugation \cite{dyson62}. Thus, three independent
parameters remain which are randomly and homogeneously taken from the
unit sphere for each SS. The phase $\varphi$ is randomly taken from
$[0,2\pi[$. In random matrix theory the symmetries (\ref{SP-UNI}) and
(\ref{SP-TR}) correspond to an ensemble of quaternion real
Hamiltonians, which can be diagonalized by symplectic transformations.
Because of this, we
also refer to this symmetry as symplectic symmetry.

We now define the {\em spin scattering strength} by
\begin{equation}
  s=\sqrt{1-q_0}=\sqrt{q_{1}^{2}+q_{2}^{2}+q_{3}^{2}}.
\label{SSstrength}
\end{equation}
This quantity takes values in the interval $[0,1]$, where $s=0$ means
no spin scattering and $s=1$ full spin scattering, resulting in full
spin relaxation after one scattering event. The parameter $s$ is fixed
for the whole network.

\subsection{Parameter space}
\begin{figure}
  \begin{center}
    \leavevmode
    \epsfxsize=5cm
    \epsffile{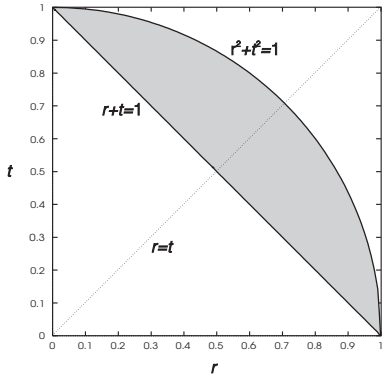}
  \end{center}
  \caption{Cross-section of the parameter space at some fixed $s$.}
  \label{PIC-Phasenraum}
\end{figure}

In conclusion there are three independent quantities building up the
three dimensional parameter space (or phase space) of the S2NC-NWM.
With the restrictions (\ref{r-t-restriction}) and $s\in[0,1]$ the
possible values $(r,t,s)$ are located in a certain volume in
${\Bbb{R}}^3$. Fig. \ref{PIC-Phasenraum} shows a cross-section
of this volume at some fixed value of $s$. The grey area contains the
allowed values. If the time reversal symmetry is omitted, i.e.
switching to the U2NC-NWM, the phase space (at some fixed s) is the
entire quarter of the circle.

There are three exceptional points in the phase space: For $r=0$ and
$t=1$ we have the delocalized fixed point, where the electron waves
propagate freely. If we take $r=1$ and $t=0$, we have the localized
fixed point, where transport stops. For $r=t=0$ the network is at the
Chalker-Coddington fixed point \cite{ChaCod88}, where only left and
right scattering exists. On the line $r^2+t^2=1$ the deflection
coefficient is zero. Thus, the system splits in independent 1D
subsystems which always show localization.  All of these properties
are independent of the value of $s$.

\begin{figure}
  \begin{center}
    \leavevmode
    \epsfxsize=7cm
    \epsffile{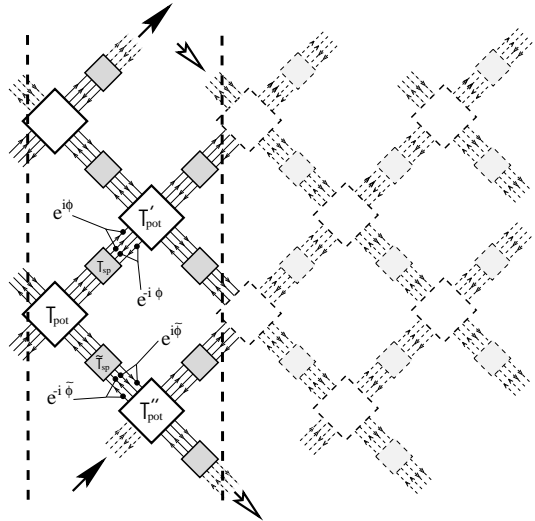}
  \end{center}
  \caption{Definition of the strip transfer matrix for $M=2$. The
    arrows indicate the periodic boundary conditions.}
  \label{PIC-T-Ausschnitt}
\end{figure}

In \cite{frech97} an elastic mean free path is defined by
\begin{equation}
  l_{\text{e}} := \frac{1}{2}\frac{t^{2}+d^{2}}{r^{2}+d^{2}}\;.
  \label{l_e}
\end{equation}
The factor $1/2$ is a consequence of the diagonal arrangement, which
is shown in Fig. \ref{PIC-T-Ausschnitt}. The unit of $l_{\text{e}}$ is a
lattice constant. Analogously, we define a spin scattering length by
\begin{equation}
  l_{\text{SO}} := \frac{1}{2}\frac{1-s^{2}}{s^{2}}\:.
\end{equation}
This length scale takes values from 0 for maximal spin scattering to
$\infty$ in the case of the absence of spin scattering.

\section{Finite-Size Scaling}
\label{FSS}
\subsection{Transfer Matrix Method}
In order to investigate the scaling behavior of the modeled system we
need some scaling variable. Although the conductance is the natural
choice for a scaling variable in this context, the {\em renormalized
  localization length} (RLL) $\Lambda=\xi/M$ is, as a self-averaging
quantity, much more convenient. Here $\xi$ is the localization length
of a quasi-1D system of width $M$ and can be calculated by the
transfer matrix method \cite{CrPaVu_Pro,PicSar81}. This method yields
a sequence of Lyapunov exponents in decreasing order, where (due to
conservation of current density) each value has a partner with
opposite sign. Beyond this, in presence of SOI each value appears
twice, because of time reversal symmetry (Kramers degeneracy). The
smallest positive Lyapunov exponent determines the quasi-1D
localization length. It is always finite due to the finite width of
the system.

In order to be able to apply the transfer matrix method we have to
convert our scattering matrices ${\sf S}_{\text{pot}}$ and ${\sf
  S}_{\text{sp}}$ into the corresponding transfer matrices ${\sf
  T}_{\text{pot}}$ and ${\sf T}_{\text{sp}}$, which map channels on
the left to channels on the right. In contrast to the scattering
matrices the transfer matrices are multiplicative, which means the
following: The total system is divided into a sequence of elementary
subsystems each of them corresponding to a single transfer step. The
transfer matrix of the total system is then given by the product of
the transfer matrices of the subsystems ("strip transfer matrices").

\subsection{Transfer Matrices in the Network Model}
With the channel orientation of Fig.  \mbox{\ref{PIC-spo}} and Fig.
\ref{PIC-spi} the transfer matrices of the S2NC-NWM are defined by
\begin{equation}
  \footnotesize
  \left ( \begin{array}{c}
    O^{+}_{3} \\ O^{-}_{3} \\ I^{+}_{3} \\ I^{-}_{3} \\ O^{+}_{4}
    \\ O^{-}_{4} \\ I^{+}_{4} \\ I^{-}_{4}
  \end{array} \right )
  = {\sf T}_{\text{pot}}
  \left ( \begin{array}{c}
    I^{+}_{1} \\ I^{-}_{1} \\ O^{+}_{1} \\ O^{-}_{1} \\ I^{+}_{2}
    \\ I^{-}_{2} \\ O^{+}_{2} \\ O^{-}_{2}
  \end{array} \right )
\end{equation}
and
\begin{equation}
  \footnotesize
  \left ( \begin{array}{c}
    \tilde O^{+} \\ \tilde O^{-} \\ \tilde I^{+} \\ \tilde I^{-}
  \end{array} \right )
  = {\sf T}_{\text{sp}}
  \left ( \begin{array}{c}
    I^{+} \\ I^{-} \\ O^{+} \\ O^{-}
  \end{array} \right )\:,
\end{equation}
respectively.  From this definition a diagonal arrangement of the
whole network results, as is shown in Fig. \ref{PIC-T-Ausschnitt}. The
arrows at the top and the bottom of the figure indicate the periodic
boundary conditions which we have chosen. The natural width unit in
this arrangement is a {\em pair} of diagonally neighbored transfer
matrices, which corresponds to a channel number of $N_{\text{c}}=8$.
Therefore the bold printed part of the picture represents a strip
transfer matrix of width $M=2$, channel number $N_{\text{c}}=16$ and
unit length $L=1$ (horizontal direction).

In the language of transfer matrices the conservation of current
density writes as {\em pseudo-unitarity}:
\begin{equation}
  {\sf T}_{\text{pot}}\Sigma_z {\sf T}_{\text{pot}}^\dagger = \Sigma_z  
\end{equation}
with
\begin{equation} 
  \Sigma_z =
  \left ( \begin{array}{cc}
    \openone_4 & 0 \\ 0 & -\openone_4
  \end{array} \right )\:,
\end{equation}
and
\begin{equation}
  {\sf T}_{\text{sp}}\Sigma_z^\prime {\sf T}_{\text{sp}}^\dagger =
  \Sigma_z^\prime
\end{equation}
with
\begin{equation}
  \Sigma_z^\prime = 
  \left ( \begin{array}{cc}
    \openone_2 & 0 \\ 0 & -\openone_2
  \end{array} \right ),
\end{equation}
respectively. On the other hand, time reversal invariance requires
\begin{equation}
  {\sf T}_{\text{pot}}=
  \left ( \begin{array}{cc}
    0 & {\sf D} \\ {\sf D} & 0
  \end{array} \right )^{\text{T}} 
  {\sf T}_{\text{pot}}^\ast
  \left ( \begin{array}{cc}
    0 & {\sf D} \\ {\sf D} & 0
  \end{array} \right )
\end{equation}
and
\begin{equation} 
  {\sf T}_{\text{sp}}= {\sf D}^{\text{T}} {\sf T}_{\text{sp}}^\ast {\sf D}\;, 
\end{equation}
with ${\sf D}$ as given in (\ref{D-DEF}). A parameterization of ${\sf
  T}_{\text{pot}}$ compatible with these restrictions is given by
\cite{frech97}
\begin{equation}
  {\sf T}_{\text{pot}} = 
  \left ( \begin{array}{cccc}
    \alpha^\ast & \gamma & \beta^\ast & \delta \\
    -\gamma & \alpha & -\delta & \beta \\
    \beta^\ast & \delta & \alpha^\ast & \gamma \\
    -\delta & \beta & -\gamma & \alpha \\
  \end{array} \right )
  \otimes \openone_2
\end{equation}
with
\begin{equation}
\begin{array}{ll}
    \displaystyle \alpha  = \displaystyle\frac{d}{\Delta}, &
    \displaystyle \beta = -\frac{t\text{e}^{\text{i}\phi_t}}{\Delta}, \\[0.5cm] 
    \displaystyle \gamma = \frac{(r\text{e}^{\text{i}{\phi_r}}-
      t\text{e}^{\text{i}\phi_t})d} {\Delta},
    & \displaystyle \delta = \frac{d^2-r{\text e}^{{\text
          i}\phi_r}t{\text e}^{{\text i}\phi_t}}{\Delta},\\[0.5cm]
    \Delta = d^2-(t{\text e}^{{\text i}\phi_t})^2.& 
  \end{array}
\end{equation}
For ${\sf T}_{\text{sp}}$ we find
\begin{equation}
  {\sf T}_{\text{sp}} =  
  \left ( \begin{array}{cc}
    {\text e}^{{\text i}{\varphi}}\bar q & 0 \\ 0 & {\text e}^{-{\text
        i}{\varphi}}\bar q
  \end{array} \right )
  \label{tsp}
\end{equation}
with $\bar q$ as in Eq. (\ref{Q-Matrix}).  We omitted the four phase
factors in ${\sf T}_{\text{pot}}$ because they can be combined with
those of ${\sf T}_{\text{sp}}$ to a resulting phase factor $\varphi$
in ${\sf T}_{\text{sp}}$.

\subsection{Localization-Delocalization Transition}
\begin{figure}
  \begin{center}
    \leavevmode
    \epsfxsize=6cm
    \epsffile{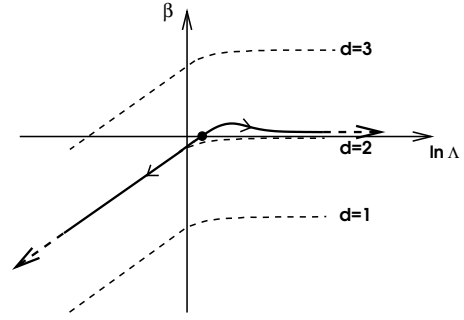}
  \end{center}
  \caption{Qualitative picture of the $\beta$-function for
    $d=1,2,3$. In the 2D case there can be a transition due to weak
    anti-localization.}
  \label{PIC-beta-funk}
\end{figure}

The scaling behavior of the RLL $\Lambda$ determines whether the
system is localized or delocalized. If $\Lambda$ shrinks with
increasing system width $M$ the system behaves like an insulator, if
it grows, the system is metallic. At the LD transition the RLL is
independent of the system width. To ensure that $\Lambda$ is in fact a
scaling variable one has to find a scaling function that is a function
of only the {\em ratio} of the correlation length $\xi_{\text{c}}$ and
$M$,
\begin{equation}
  \Lambda(M) = \tilde f\biggl(\frac{\xi_{\text{c}}}{M}\biggr),
\label{Lambda-Skalenf-tilde}
\end{equation}
or logarithmically
\begin{equation}
  \ln\Lambda(M) = f\Bigl(\ln M - \ln \xi_{\text{c}}\Bigr).
\label{Lambda-Skalenf}
\end{equation}
Equivalent to the existence of a scaling function is the formulation
of a flow equation with a $\beta$-function that is a function of
$\ln\Lambda$ only,
\begin{equation}
  \beta(\ln \Lambda) = \frac{\text{d}\ln \Lambda}{\text{d}\ln M}.
\end{equation}
Fig. \ref{PIC-beta-funk} shows a qualitative picture of the
$\beta$-function in different dimensions. Due to Ohm's law the
limiting value of $\beta$ for large conductance (or $\Lambda$) is
$d-2$. Without SOI there is a negative correction to the conductance
caused by weak localization \cite{berg84}. The $\beta$-function is
always negative in 2D and therefore all states are localized. In the
presence of SOI the correction to the metallic conductance changes the
sign due to weak anti-localization \cite{berg82,HiLaNa80,MaeFuk81}.
Consequently, in 2D and in the large conductance limit the
$\beta$-function converges to zero from above. Since in the strongly
disordered regime all states are exponentially localized, the
existence of a LD transition in the 2D symplectic case follows from
simple scaling arguments. At the critical point, where the
$\beta$-function is zero, the correlation length shows power low
scaling with the critical exponent $\nu$,
\begin{equation}
  \xi_{\text{c}}(r) = \xi_{\text{c}}^0\left | r-r^\ast \right |^{-\nu},
\label{xic-r}
\end{equation}
where $r$ is a system parameter, e.g. the reflection coefficient of
the NWM, and $r^\ast$ is its critical value. Here the correlation
length is the fictitious system width up to which the system is in the
critical regime. In the localized regime this is just the quasi-1D
localization length obtained by the transfer matrix method in the
thermodynamic limit,
\begin{equation}
  \xi_{\text{c}} = \lim_{M\rightarrow\infty} \xi(M) = \xi_{\infty}.
\end{equation}

In \cite{jans94} it was shown that the {\em typical local density of
  states} (LDOS) is an appropriate choice for the order parameter of
the LD transition,
\begin{equation}
  \rho_{\text{typ}} := {\text e}^{\left\langle \ln \rho(E,\bbox{r})
    \right\rangle} \propto \left | r-r^\ast \right |^{\beta_{\rho}},
\end{equation}
with $\rho(\bbox{r}) = \left | \psi(E,\bbox{r}) \right |
^2/\Delta(E)$, energy $E$ and local level spacing $\Delta(E)$.
$\left\langle\ldots\right\rangle$ denotes disorder average and
$\beta_\rho$ is the critical exponent of the order parameter.
Furthermore, $\rho_{\text{typ}}$ shows power law scaling at the
critical point with exponent $d-\alpha_0$,
\begin{equation}
  \rho_{\text{typ}} \propto L^{d-\alpha_0},
\end{equation}
where $L^d$ is the volume of a $d$-dimensional cube and $\alpha_0$ is
a scaling exponent, which is known from multifractal analysis of
critical wave functions. In 2D, i.e. for a square system, this scaling
exponent is linked to the critical value of the quasi-1D RLL by a
conformal mapping argument \cite{jans94},
\begin{equation}
  \Lambda^\ast = \frac{1}{\pi(\alpha_0-2)}.
\label{alpha_null}
\end{equation}
With the knowledge of $\nu$ and $\alpha_0$ the critical exponent of
the LDOS is given by
\begin{equation}
  \beta_\rho = \nu(\alpha_0-2).
\end{equation}

\section{Methods of evaluation}
\label{MOE}
\subsection{Fit Procedure for the Scaling Function}
\label{ScalF-Fit}

In our work we follow the method introduced in \cite{hucke90}, which
not only fits the scaling function to the data of the RLLs but also
uses a $\chi^2$ test to check the confidence of the fit.

According to (\ref{Lambda-Skalenf}) we want to fit the scaling
function $f$ to the logarithms of the RLLs, which depend on the $n_M$
system widths $\{M_1,\ldots,M_{n_M}\}$ and the $n_r$ system parameters
$\{r_1,\ldots,r_{n_r}\}$. So we have $n_\Lambda = n_M\cdot n_r$ data
points $\{\Lambda_1,\ldots,\Lambda_{n_\Lambda}\}$. For abbreviation we
now introduce the following vectors,
\begin{equation}
  \bbox{Y} := 
  \left ( \begin{array}{c}
    \ln \Lambda_1 \\ \vdots \\ \ln \Lambda_{n_\Lambda}
  \end{array} \right ),
  \qquad
  \bbox{\tilde X} := 
  \left ( \begin{array}{c}
    \ln M_1 - \ln \xi_c(r_1) \\ \vdots \\  \ln M_{n_M} - \ln
    \xi_c(r_{n_r})
  \end{array} \right )
\end{equation}
and
\begin{equation}
  \bbox{E} := 
  \left ( \begin{array}{c}
    \Delta \ln \Lambda_1 \\ \vdots \\ \Delta\ln \Lambda_{n_\Lambda}
  \end{array} \right ),
\end{equation}
the latter being the vector containing the errors of the average
obtained by the transfer matrix method. Assuming the data to be
statistically independent the corresponding correlation matrix
${\sf C}_\Lambda$ given by
\begin{equation}
  \left({\sf C}_\Lambda\right)_{ij} := \langle E_i \cdot E_j \rangle
\end{equation}
is diagonal. Here $\langle\dots\rangle$ denotes the mean value.

Following the procedure in Ref. \cite{hucke90} we make an ansatz for
the scaling function by a linear combination of Chebyshev polynomials,
\begin{equation}
  F(x_i) = \frac{c_0}{2} + \sum_{k=1}^N c_k T_k(X_i),
\end{equation}
which gives a polynomial of degree $N$. Omitting the tilde over the
$X_i$ indicates the argument being rescaled to $[-1,1]$. In this
interval the Chebyshev polynomials are orthogonal and have simple
behavior at the edges.

The smallest of the parameters $\ln\xi_c$ is fixed by requiring
$\ln\xi_c(r_1) \equiv 0$ for the delocalized branch and
$\ln\xi_c(r_{n_r}) \equiv 0$ for the localized branch, respectively.
So we have the $n_\Theta = n_r + N$ parameters
\begin{equation}
  \bbox{\Theta} := 
  \left ( \begin{array}{c}
    \ln \xi_c(r_1) \\ \vdots \\ \ln \xi_c(r_{n_r-1}) \\ c_0 \\ \vdots
    \\ c_N
  \end{array} \right ),
\end{equation}
for the localized and
\begin{equation}
  \bbox{\Theta} := 
  \left ( \begin{array}{c}
    \ln \xi_c(r_2) \\ \vdots \\ \ln \xi_c(r_{n_r}) \\ c_0 \\ \vdots \\
    c_N
  \end{array} \right ),
\end{equation}
for the delocalized branch, respectively. These parameters have to be
fitted with respect to the data $\bbox{Y}$. Hence, the $n_\Lambda$
values of the fit function can be written as
\begin{equation}
  \bbox{F}(\bbox{X};\bbox{\Theta}) := 
  \left ( \begin{array}{c}
    F(X_1;\bbox{\Theta}) \\ \vdots \\ F(X_{n_\Lambda};\bbox{\Theta}) 
  \end{array} \right ). 
\end{equation}

We use the method of the least-squares fit, i.e. we have to minimize
the sum $S_\Theta$ of the weighted quadratic deviations, which means
solving
\begin{equation}
  \frac{\partial S_{\Theta}}{\partial \bbox{\Theta}} = \bbox{0}
\end{equation}
with
\begin{equation}
  S_{\Theta} =
  \Bigl(\bbox{Y}-\bbox{F}(\bbox{X};\bbox{\Theta})\Bigr)^{\text{T}}
  {\sf C}_\Lambda^{-1}
  \Bigl(\bbox{Y}-\bbox{F}(\bbox{X};\bbox{\Theta})\Bigr).
\end{equation}

Since $\bbox{F}$ is non-linear in the parameters $\bbox{\Theta}$ this
procedure leads to a system of $n_\Theta$ coupled non-linear
equations. Therefore $S_\Theta$ is minimized directly by a numerical
method. Starting with some estimated initial values for the logarithms
of the correlation lengths we successively optimize the $c_i$ and the
$\ln\xi_c(r_i)$. If the data are compatible we will have convergence,
and thus we can stop when a chosen accuracy is reached. The result
then is a set of coefficients $c_i$ which defines the fit function and
a set of optimized values for the correlation lengths.

According to Ref. \cite{mart_Stat} the correlation matrix of the
parameters is given by
\begin{equation}
  {\sf C}_\Theta = \left({\sf F}_\Theta^{\text{T}}{\sf
      C}_\Lambda^{\text{-1}} {\sf F}_\Theta\right)^{-1}, 
\label{C-Theta}
\end{equation}
where ${\sf F}_\Theta$ is the Jacobian of $\bbox{F}$ with respect to
$\bbox{\Theta}$,
\begin{equation}
  {\sf F}_\Theta := \frac{\partial\bbox{F}}{\partial\bbox{\Theta}}.
\end{equation}
Usually, as errors of the parameters one takes the diagonal elements
of the error matrix ${\sf E}$, which is defined by
\begin{equation}
  {\sf E}_\Theta = \frac{S_\Theta}{n_\Lambda-n_\Theta}{\sf C}_\Theta.
\end{equation}

\subsection{Testing the Fit of the Scaling Function}
A converging fit procedure doesn't guarantee that the errors of the
numerical data are actually compatible with the obtained fit function.
Therefore, in addition to this fit procedure we apply a $\chi^2$ test
to estimate the confidence of the fit. We make the essential
assumption that the data $y_i$ are {\em normally} distributed about
$f(x_i;\bbox{\Theta})$ with variances ${e_i}^2$.  Consequently the
quantity $S_\Theta$ has to be distributed as $\chi^2$ with
$n_\Lambda-n_\Theta$ degrees of freedom. This distribution has the
estimated value $n_\Lambda-n_\Theta$ and the variance
$2(n_\Lambda-n_\Theta)$. A suitable measure for the confidence of the
fit then is the normalized deviation of $S$ from the estimated value
\begin{equation}
  \Delta_{\Theta} =
  \frac{S_{\Theta}-(n_\Lambda-n_\Theta)}{\sqrt{2(n_\Lambda-n_\Theta)}}.
  \label{Delta-Theta}
\end{equation}
If $\left | \Delta_\Theta \right |\lesssim 1$ it is safe to assume
that the fit is trustworthy. But if $\left | \Delta_\Theta \right |$
takes values much larger than 1 it is very unlikely that the data
$Y_i$ are normally distributed about $F(X_i;\Theta)$ which indicates
systematic errors.  In this case the fit has failed, we have to give
up the assumption of one-parameter scaling and further calculations,
e.g. of the critical exponent, don't make much sense.

It should be noticed that rescaling of the variance matrix ${\sf S}$
by a factor $b$, $\tilde{\sf S} = b\cdot{\sf S}$, results in the
reciprocal rescaling of $S_\Theta$, $\tilde S_\Theta =
b^{-1}S_\Theta$. Since $\Delta_\Theta$ depends sensitively on
$S_\Theta$, especially if $n_\Lambda$ is small, one should carefully
consider, how to determine the errors of the raw data.

\subsection{Fitting the Critical Exponent $\bbox{\nu}$ and
  $\bbox{r}^\ast$\\ and Testing the Fit}
\label{nu-Fit}
In order to determine the critical exponent of the correlation length
we apply the same idea as before with respect to Eq. (\ref{xic-r}).
Particularly, we simultaneously deal with both branches of the scaling
function, i.e. we assume the critical value $r^\ast$ and the critical
exponent $\nu$ to be the same in the localized and the delocalized
regime. Only the prefactor can take different values, which will be
denoted as $\xi_{\text{c}}^{\text{0,loc}}$ and
$\xi_{\text{c}}^{\text{0,deloc}}$ for the localized and the
delocalized regime, respectively. Taking the logarithm Eq.
(\ref{xic-r}) writes as
\begin{mathletters}
  \label{ln-r}
  \begin{equation}
    \ln \xi_{\text{c}}(r_i)  = \ln
    \xi_{\text{c}}^{\text{0,loc}} - \nu \ln \left |
      r_i-r^\ast \right |
  \end{equation}
  and
  \begin{equation}
    \ln \xi_{\text{c}}(r_i)  = \ln \xi_{\text{c}}^{\text{0,deloc}} - \nu 
    \ln \left | r_i-r^\ast \right |
  \end{equation}
\end{mathletters}
for $r_i$ in the localized and delocalized regime, respectively.

The $n_r$ values $\ln \xi_{\text{c}}(r_i)$ are the results of the
foregoing optimization, the arguments are the values $\ln \left |
  r_i-r^\ast \right |$ and the four parameters that have to be
optimized are $\xi_{\text{c}}^{\text{0,loc}}$,
$\xi_{\text{c}}^{\text{0,deloc}}$, $\nu$ and $r^\ast$. Introducing the
vectors
\begin{mathletters}
  \begin{equation}
    \bbox{y} := 
    \left ( \begin{array}{c}
        \ln \xi_{\text{c}}(r_1) \\
        \vdots \\
        \ln \xi_{\text{c}}(r_{n_r})
      \end{array} \right ),
    \quad
    \bbox{x} := 
    \left ( \begin{array}{c}
        \ln \left | r_1-r^\ast \right | \\
        \vdots \\
        \ln \left | r_{n_r}-r^\ast \right |
      \end{array} \right )
  \end{equation}
  \begin{equation}
    \bbox{\theta} :=
    \left ( \begin{array}{c}
        \ln \xi_{\text{c}}^{\text{0,loc}} \\
        \ln \xi_{\text{c}}^{\text{0,deloc}} \\
        \nu \\
        r^\ast
      \end{array} \right )
    \quad \text{and} \quad
    \bbox{\tilde \theta} := 
    \left ( \begin{array}{c}
        \ln \xi_{\text{c}}^{\text{0,loc}} \\
        \ln \xi_{\text{c}}^{\text{0,deloc}} \\
        \nu
      \end{array} \right )
  \end{equation}
\end{mathletters}
and comparing them with Eq. (\ref{ln-r}) the fit function can be
written as
\begin{equation}
  \bbox{f}(\bbox{x};\bbox{\theta}) =
  {\sf W}(\bbox{x})\bbox{\tilde\theta},
\end{equation}
with
\begin{equation}
  {\sf W} :=
  \left ( \begin{array}{ccc}
    1 & 0 & - \ln \left | r_1-r^\ast \right | \\
    \vdots & \vdots & \vdots\\
    1 & 0 & - \ln \left | r_{n_r{_{\text{,loc}}}}-r^\ast \right | \\
    0 & 1 & - \ln \left | r_{n_r{_{\text{,loc}}}+1}-r^\ast \right | \\
    \vdots & \vdots & \vdots\\
    0 & 1 & - \ln \left | r_{n_r}-r^\ast \right |
  \end{array} \right ).
\end{equation}
Here $n_{\text{r,loc}}$ is the number of values $r_i$ which belong to
the localized regime.

The correlation matrix ${\sf C}_{\xi_{\text{c}}}$ of the data $y_i$ is
the upper left $n_r\times n_r$ submatrix of ${\sf C}_\Theta$ (Eq.
(\ref{C-Theta})) obtained by the fit of the scaling function. Thus,
the sum of the quadratic deviations is
\begin{equation}
  S_\theta =
  \Bigl(\bbox{y}-\bbox{f}(\bbox{x};\bbox{\theta})\Bigr)^{\text{T}}
  {\sf C}_{\xi_{\text{c}}}^{-1}
  \Bigl(\bbox{y}-\bbox{f}(\bbox{x};\bbox{\theta})\Bigr).
\end{equation}
Since the fit function is a {\em linear} function in the argument
$\bbox{x}$, we can analytically solve the minimization problem
\begin{equation}
  \frac{\partial S_\theta}{\partial \bbox{\tilde\theta}} = \bbox{0},
\end{equation}
which leads to
\begin{equation}
  \bbox{\tilde\theta} = \left({\sf W}^{\text{T}}{\sf
      C}_{\xi_{\text{c}}}^{-1} {\sf W}\right)^{-1} {\sf
    W}^{\text{T}}{\sf C}_{\xi_{\text{c}}}^{-1}
  \bbox{y}.
\end{equation}
In contrast to that the parameter $r^\ast$ has to be optimized
numerically since it appears non-linearly in Eq. (\ref{ln-r}). Giving
some starting value for $r^\ast$ one iteration step consists of
successively optimizing $\bbox{\tilde\theta}$ and $r^\ast$.
  
The $4\times4$ correlation matrix of the four parameters is given by
\begin{equation}
  {\sf C}_{\theta} =
  \left({\sf F}_{\theta}^{\text{T}}{\sf C}_{\xi_{\text{c}}}^{-1}
    {\sf F}_{\theta}\right)^{-1},
\end{equation}
where 
\begin{equation}
  {\sf F}_{\theta} := \frac{\partial\bbox{f}}{\partial\bbox{\theta}}.
\end{equation}
Finally we get the error matrix
\begin{equation}
  {\sf E}_\theta = \frac{S_\theta}{n_r-4}{\sf C}_\theta.
\label{E_theta}
\end{equation}

As in Eq. (\ref{Delta-Theta}) we use the quantity
\begin{equation}
  \Delta_{\theta} = \frac{S_{\theta}-(n_r-4)}{\sqrt{2(n_r-4)}}.
  \label{Delta-theta}
\end{equation}
to test the confidence of the fit. $\left | \Delta_\theta \right |$
should take values of about 1 or smaller to verify the assumption that
the values $y_i$ are normally distributed about the fit function,
which is a straight line in this case. Note that it is important to
take the correlations between the $\xi_{\text{c}}(r_i)$ introduced by
the previous fit procedure into account in the present analysis. Thus,
a simple linear regression will not give the correct results.

\subsection{Determination of $\bbox{\Lambda}^\ast$}
The fit procedures introduced in Sects. \ref{ScalF-Fit} and
\ref{nu-Fit} are not the most direct way to determine $\nu$ from the
raw data. Instead, one can fit the RLLs as functions of $\left |
  r-r^\ast \right |$ with a width dependent scaling factor $M^{1/\nu}$
\begin{equation}
  \Lambda(M;r) = h\left(M^{1/\nu}\left | r-r^\ast \right |\right),
  \label{Lambda-non-log}
\end{equation}
which is a consequence of Eqs. (\ref{Lambda-Skalenf-tilde}) and
(\ref{xic-r}). This procedure leads to a continuous curve, because
there is no splitting in two branches as in the logarithmic case.
Fitting the scaling function in this manner allows a direct evaluation
of $\Lambda^\ast$. Since at $r=r^\ast$ the argument of $h$ is zero,
one only has to calculate the function value:
\begin{equation}
  \Lambda^\ast = h(0).
  \label{Lambda-krit}
\end{equation}
Following the law of propagation of errors we get for the error
\begin{equation}
  \Delta \Lambda^\ast = \Delta \Lambda\bigr|_{r={r^\ast}} =
  M^{1/\nu}\Delta h^\prime(0) r^\ast,
\end{equation}
where $h^\prime$ denotes the first derivative of $h$ with respect to
the argument.

\section{Results and Discussions}
\label{RAD}
\subsection{Localization Lengths}
We calculated quasi-1D localization lengths for system lengths up to
$L=2\cdot10^5$ and widths from $M=2$ up to $M=32$, which corresponds
to channel numbers from $N_{\text{c}}=16$ to $N_{\text{c}}=256$.  The
corresponding errors $\Delta\Lambda$ are the errors of the average,
which vanish for $L\rightarrow\infty$. We have chosen such system
lengths $L$ that the relative errors $\Delta\Lambda / \Lambda$ take
values of about 0.1 \% to 1 \%.

\begin{figure}
  \begin{center}
    \leavevmode
    \epsfxsize=7cm
    \epsffile{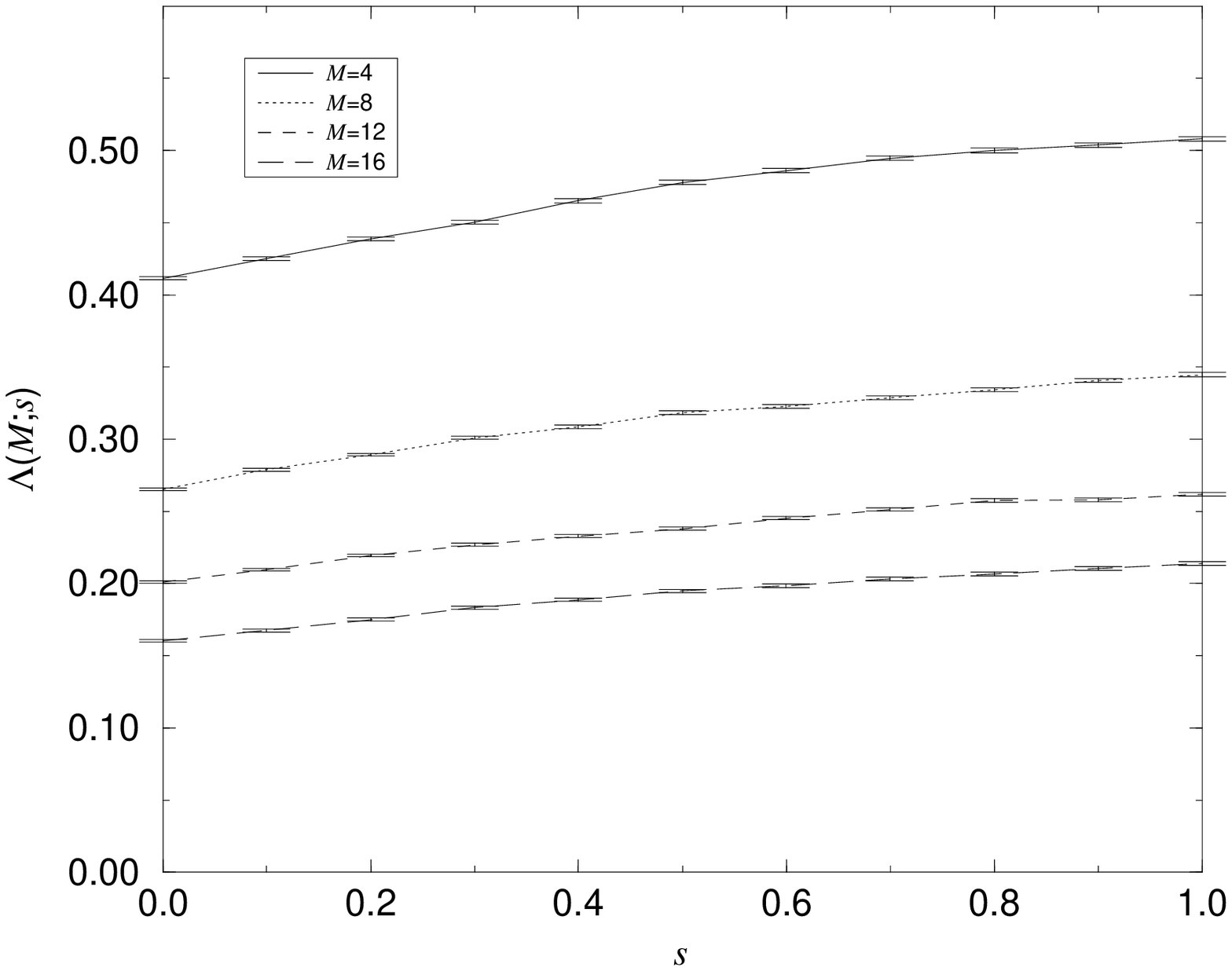}
  \end{center}
  \caption{Renormalized localization length for strong disorder, 
    $r=0.8$, $t=0.4$ which corresponds to $l_{\text{e}}\simeq0.35$,
    depending on the spin scattering strength $s$ and the the system
    width $M$.}
  \label{r8-t4-LLs}
\end{figure}

Fig. \ref{r8-t4-LLs} shows the renormalized localization lengths for
$r=0.8$ and $t=0.4$ decreasing with increasing system width for the
whole range of the spin scattering strength. So the system is in the
deeply localized regime. This matches with the fact that the mean free
path (see Eq. (\ref{l_e})) takes a value of $l_{\text{e}}\simeq0.35$
in units of the lattice constant. So the reflection is too strong to
allow extended states at all.

\begin{figure}
  \begin{center}
    \leavevmode
    \epsfxsize=7cm
    \epsffile{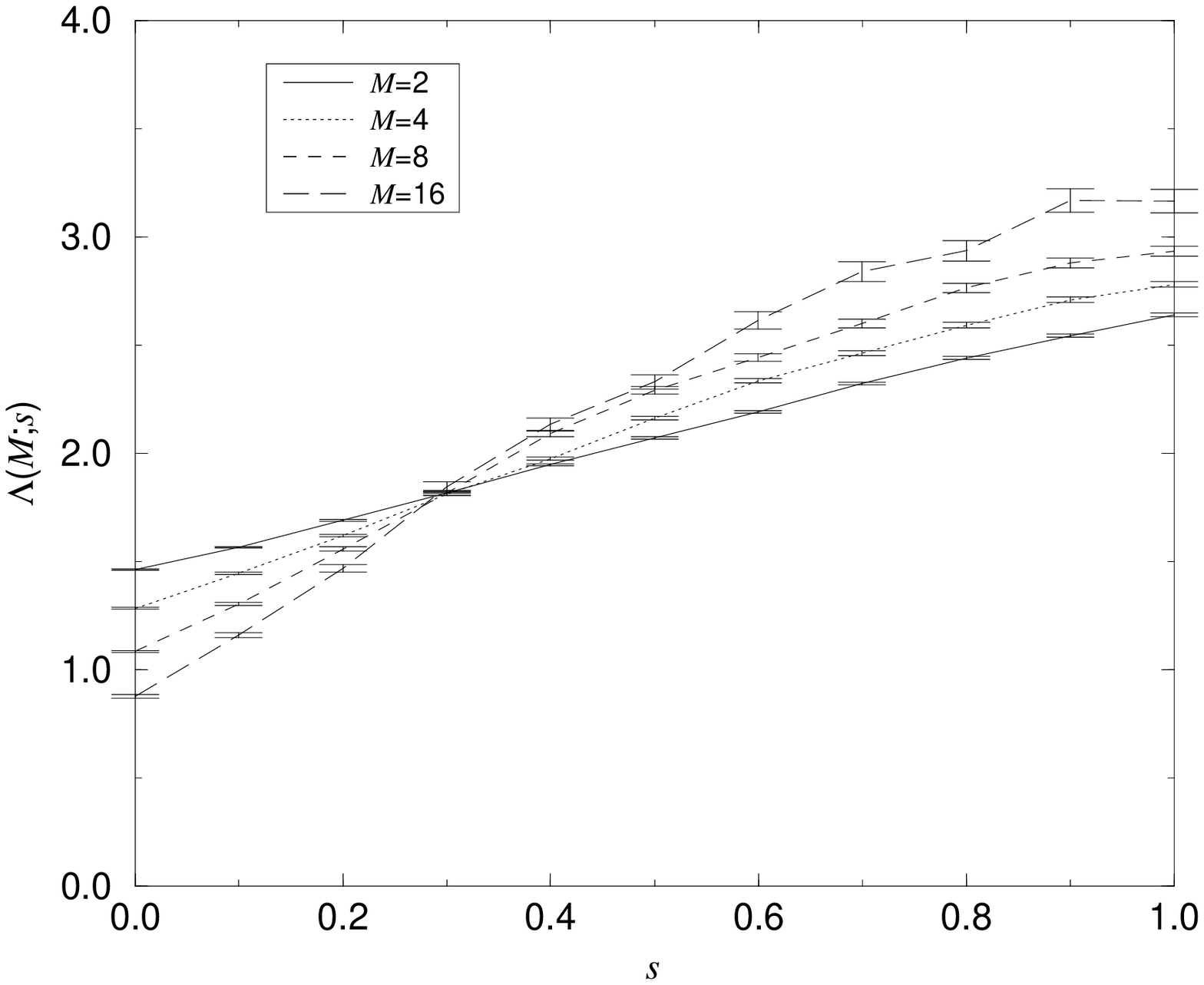}
  \end{center}
  \caption{Renormalized localization length for $r=0.55$, $t=0.6$
    which corresponds to $l_{\text{e}}\simeq1.1$, depending on the spin
    scattering strength $s$ and the the system width $M$.}
  \label{r55-t6-LLs}
\end{figure}

Contrary to this, the next example exhibits a LD transition at
$s\simeq0.3$, which can be seen in Fig. \ref{r55-t6-LLs}. With
$r=0.55$ and $t=0.6$ corresponding to $l_{\text{e}}\simeq1.1$ the
effect of weak anti-localization causes the existence of extended
states, if $s$ is strong enough. The intersection of the curves
clearly indicates the LD transition. However, their slope is rather
small compared to the error bars preventing an accurate scaling
analysis close to the critical point.

\begin{figure}
  \begin{center}
    \leavevmode
    \epsfxsize=7cm
    \epsffile{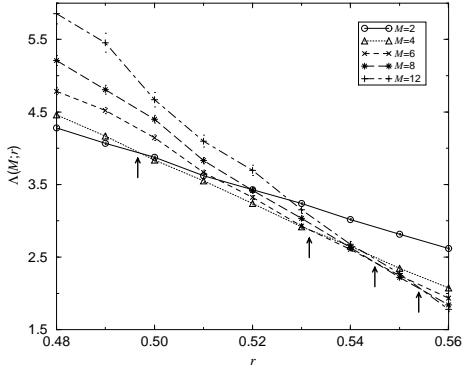}
  \end{center}
  \caption{Renormalized localization length in the vicinity of the
    localization-delocalization transition, $t=0.8$, $s=0.6$ which
    corresponds to $l_{\text{e}}\simeq2$, depending on the reflection
    coefficient $r$ and the the system width $M$. The arrows mark the
    section of curves corresponding to neighbored values of $M$.}
  \label{t8-s6-LLs}
\end{figure}

The third example shown in Fig. \ref{t8-s6-LLs} belongs to the values
$t=0.8$ and $s=0.4$, where $r$ varies from $r=0.48$ to $r=0.56$. It
exhibits another kind of problem. Although curves for different system
widths $M$ intersect, the points of intersection systematically depend
on the width. The larger $M$ is the closer are the points of
intersection for curves of neighboring values of $M$. It is obvious
that there exists a limiting point, which would be the true critical
point. The observed deviations are the consequence of finite-size
effects, which vanish in the thermodynamic limit. By investigating
$\Lambda$ at a lot of different points in parameter space we have seen
that the deviations due to finite-size effects are larger for more
delocalized systems. Actually, there is only a small area in parameter
space that is suitable for a quantitative analysis of the LD
transition.

\subsection{Phase Diagram}
\begin{figure}
    \hspace*{1cm}\begin{tabular}{c@{\hspace{0.4cm}}c}
      \leavevmode
      \epsfysize=3.1cm
      \epsffile[121 198 531 603]{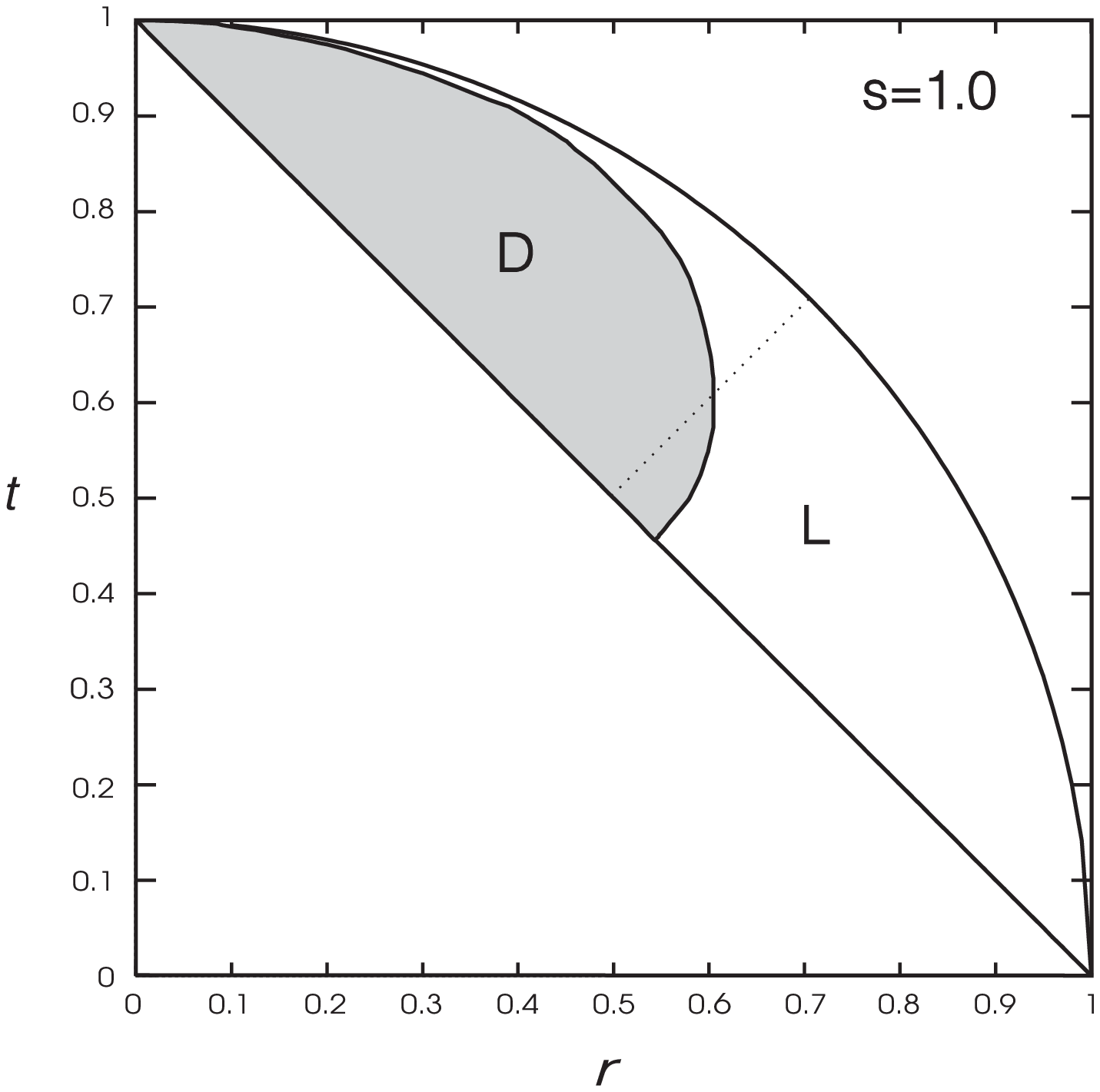}&
      \epsfysize=3.1cm
      \epsffile[118 356 537 767]{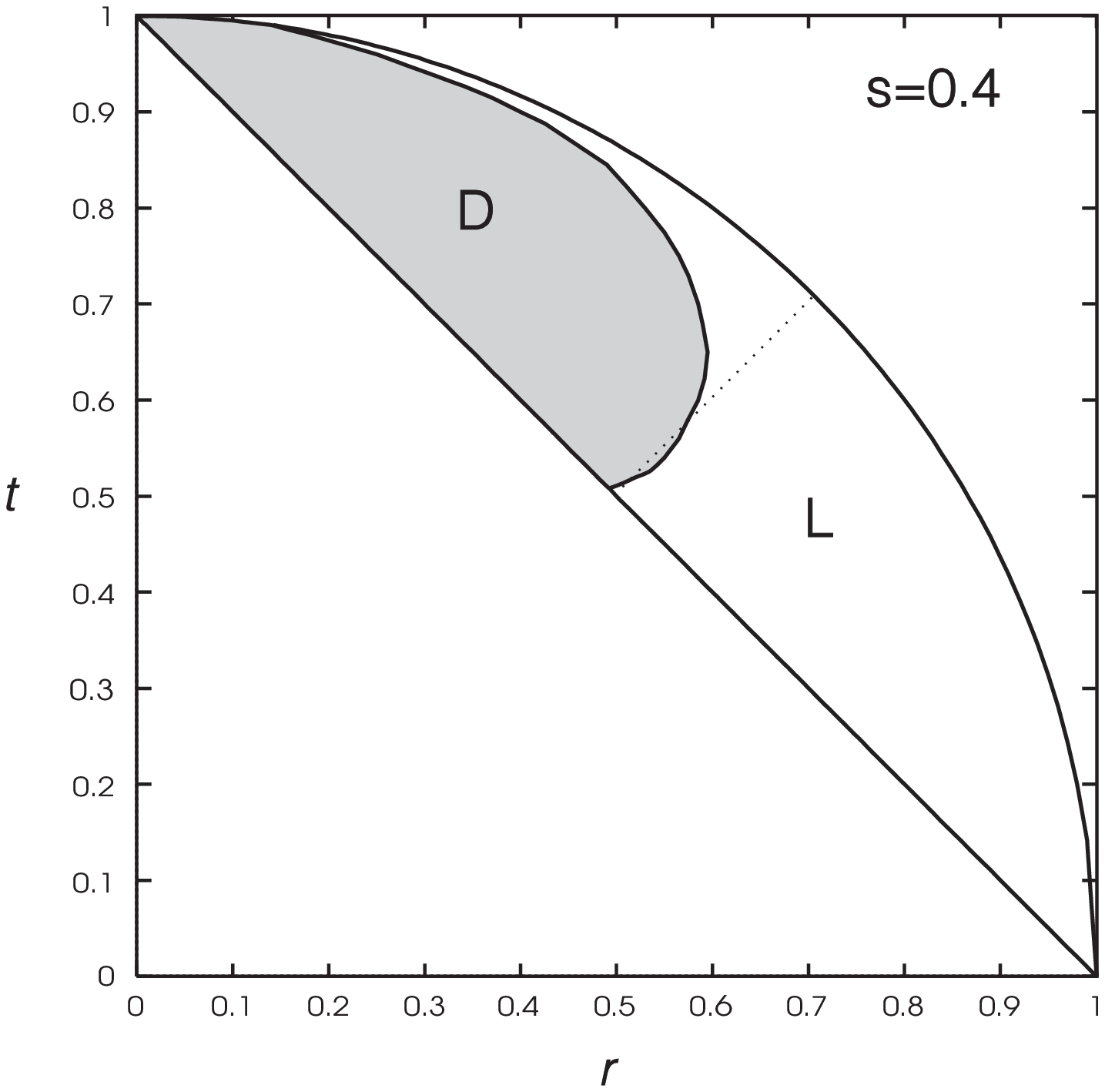}\\
      \epsfysize=3.1cm
      \epsffile[61 199 472 604]{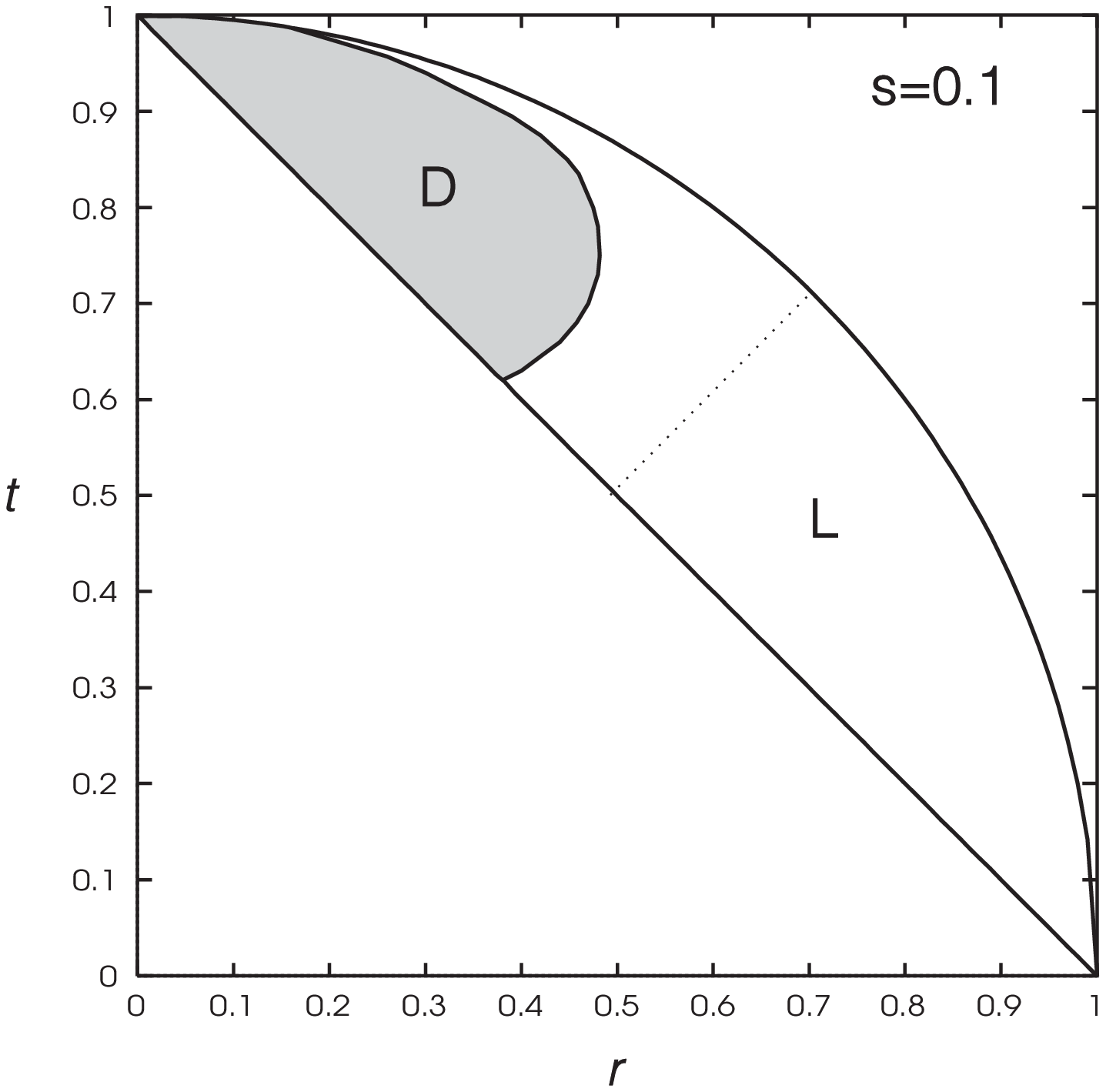}&
      \epsfysize=3.1cm
      \epsffile[119 196 531 600]{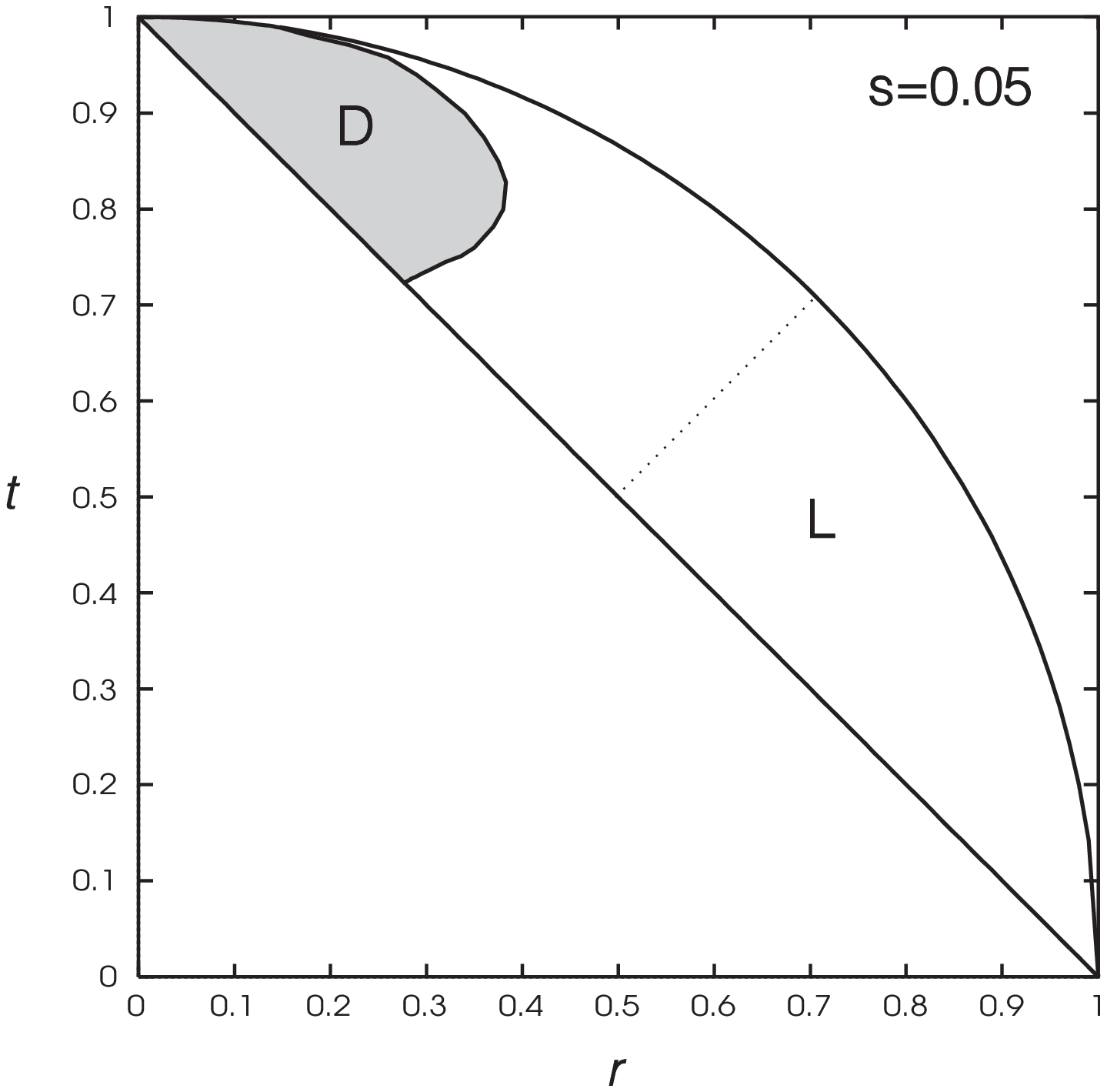}\\
      \epsfysize=3.1cm
      \epsffile[119 198 529 600]{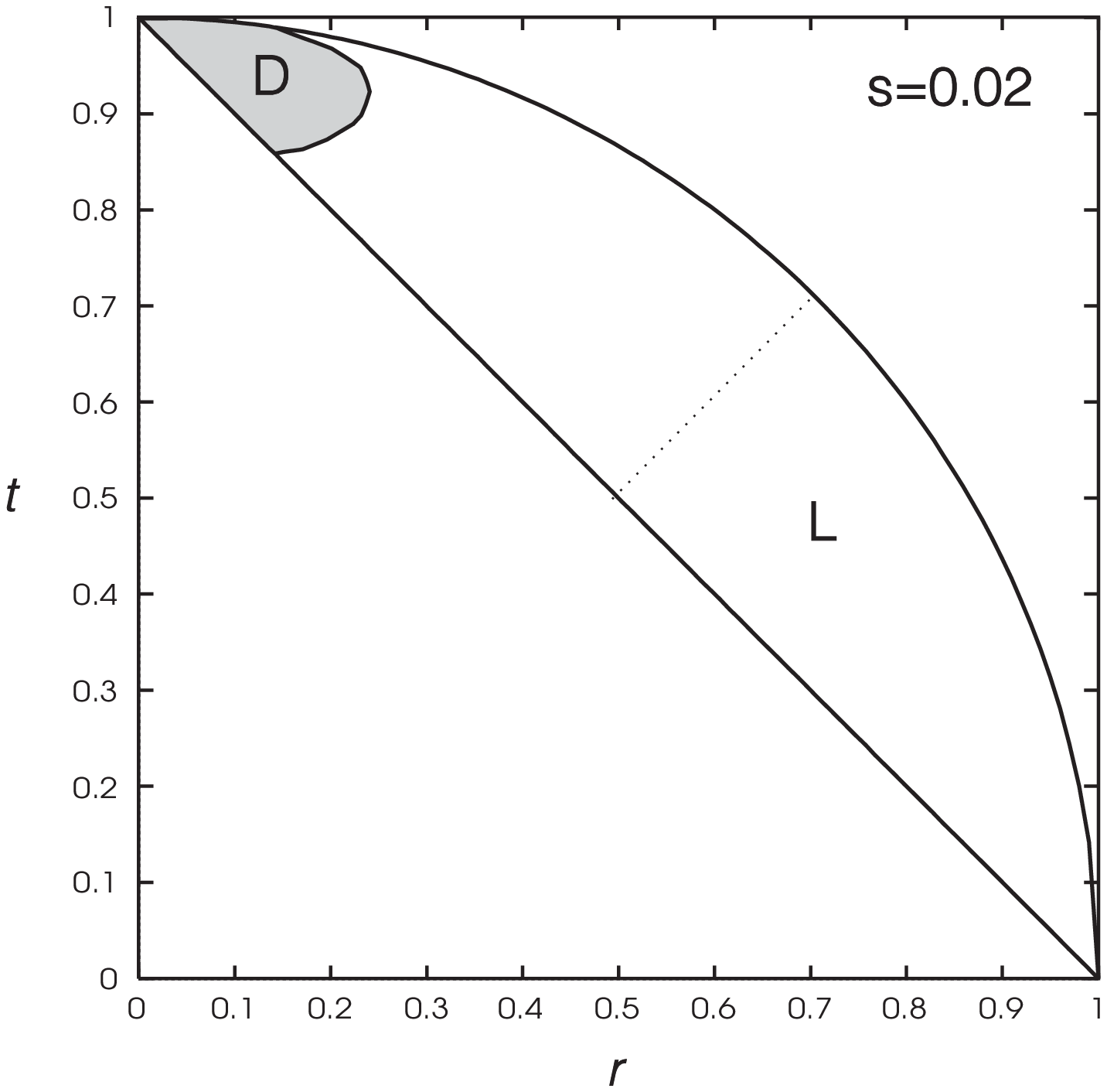}&
      \epsfysize=3.1cm
      \epsffile[119 195 530 600]{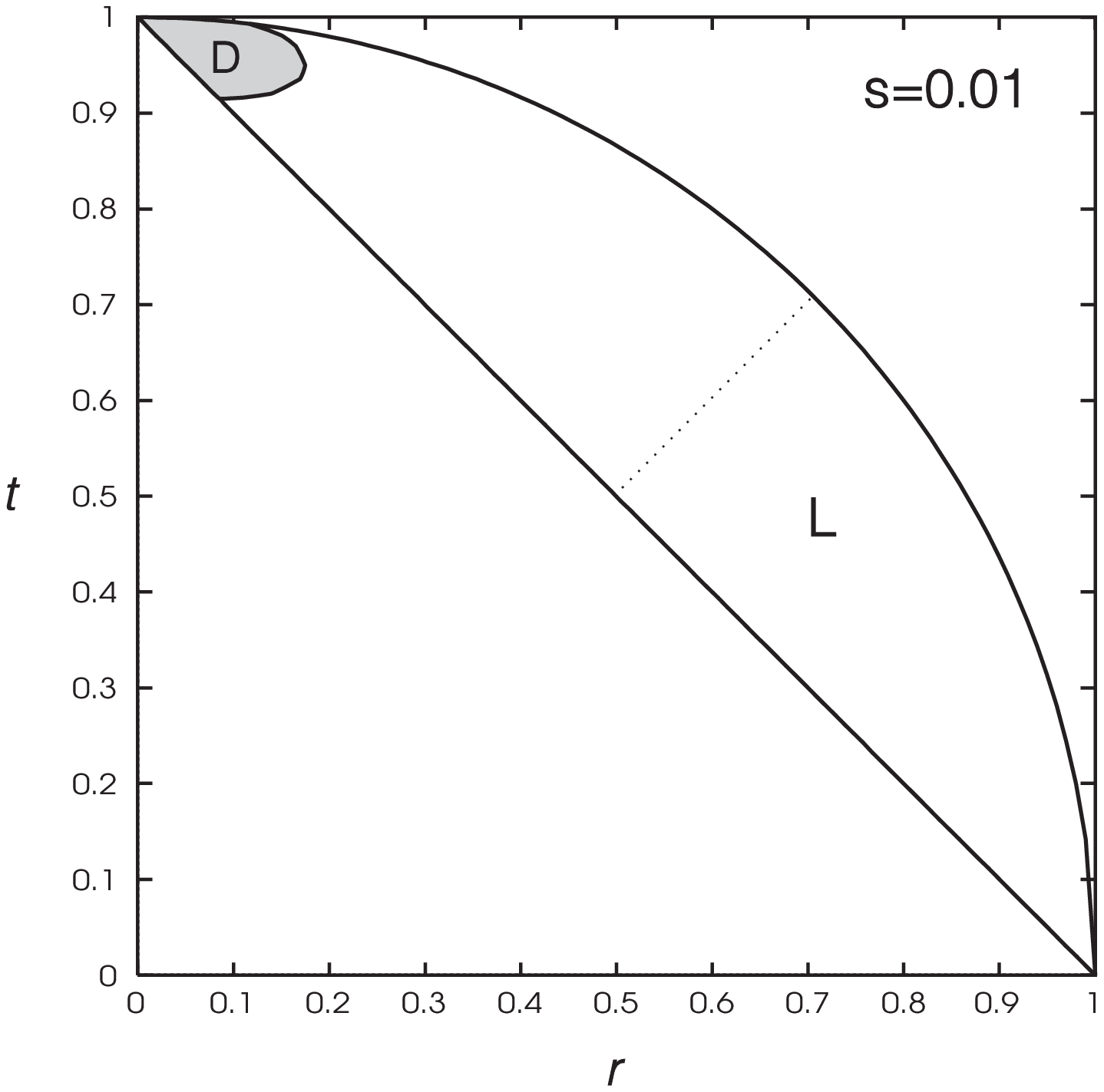}
  \end{tabular}
  \caption{Phase diagrams for the localization-delocalization transition 
    for cross-sections at $s=$ 1, 0.4, 0.1, 0.05, 0.02, 0.01. The grey
    area shows the delocalized, the white the localized regime. The
    dotted line corresponds to $r=t$.}
  \label{Phasendiagramm}
\end{figure}

We determined a phase diagram for the LD transition by using the
scaling behavior of the RLL. More precisely, we calculated
$\Lambda(M_1=4)$ and $\Lambda(M_2=8)$ for a lot of pairs $(r,t)$ with
$s$ = 0.01, 0.02, 0.05, 0.1, 0.4 and 1. In order to get the critical
line in the $(r,t)$-subspace with fixed $s$ we decided the point
$(r,t,s)$ to be in the localized and delocalized regime, if
\begin{equation}
  \Lambda(M_{1})-\Delta\Lambda(M_{1}) \;>\;
  \Lambda(M_{2})+\Delta\Lambda(M_{2})
\end{equation}
and
\begin{equation}
  \Lambda(M_{2})-\Delta\Lambda(M_{2}) \;>\;
  \Lambda(M_{1})+\Delta\Lambda(M_{1}),
\end{equation}
respectively. In the case that both conditions failed, we considered
the point in the parameter space to be critical. By that procedure we
got a critical {\em region}, i.e. the separating line had some finite
width.  Fig.  \ref{Phasendiagramm} shows the resulting phase diagrams
for some intersections of the parameter space at the above declared
fixed values of $s$. The white area marks the localized the grey one
the delocalized phase. The drawn critical line is the optical
interpolated center line of the critical region.

The region of the metallic phase shrinks with decreasing spin
scattering strength. This is due to the fact that the weak
anti-localization then becomes less effective in preventing Anderson
localization. At $s=0$ the system changes universality class from
symplectic to orthogonal symmetry, a fact that could be verified by
comparing the values of the localization lengths with those in Ref.
\cite{frech97}. On the other hand even a very small value of $s$ gives
rise to a certain delocalized phase, if $r$ is small and $t$ is large
enough. Of course, the larger $l_e$, i.e. the larger $t$ and smaller
$r$, the more easily extended states occur. But even in the presence
of full spin scattering only about the half of the area of the
parameter space belongs to the metallic phase. This is due to the fact
that parameter values $(r,t)$ belonging to the localized phase
correspond to too strong disorder resulting in too strong localization
to be broken by weak anti-localization.

It should be noticed, that for $t\gtrsim0.6$ the shape of the phase
boundary is influenced by finite-size effects. So the phase diagram
can only serve as a qualitative picture of the LD transition. In order
to improve on the phase diagram, one has to consider much larger
systems, which is very computer time consuming taking such a large
number of data points into account. Nevertheless, the lower part of
the boundary, i.e. the region close to the line $r=t$ (dotted line in
the figure), is suitable for quantitative investigations, as will be
shown in the following.

\subsection{Scaling Function}
\begin{figure}[tb]
  \begin{center}
    \leavevmode
    \epsfxsize=7cm
    \epsffile{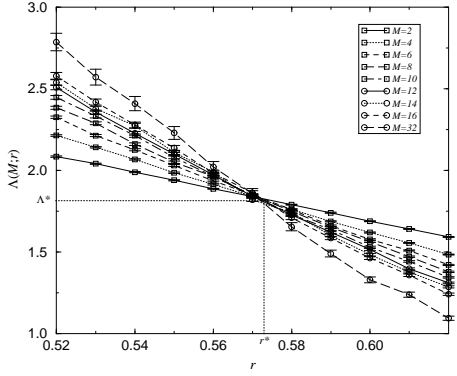}
  \end{center}
  \caption{Renormalized localization length for $t=0.6$, $s=0.4$ which
    corresponds to $l_{\text{e}}\simeq1$, depending on the reflection
    coefficient $r$ and the the system width $M$.}
\label{t6-s4-LLs}
\end{figure}

We determined the scaling function by the fitting procedure described
in Sec. \ref{ScalF-Fit} for $t=0.6$, $s=0.4$ and $r\in [0.52,0.62]$.
In this small region of the phase space the corresponding curves
$\Lambda(M;r)$ (see Fig. \ref{t6-s4-LLs}) are very well suitable for a
quantitative analysis, because of their strong $r$ dependence and the
absence of noticeable finite-size effects ($l_{\text{e}}\simeq1$).

\begin{figure}[tb]
  \begin{center}
    \leavevmode
    \epsfxsize=7cm
    \epsffile{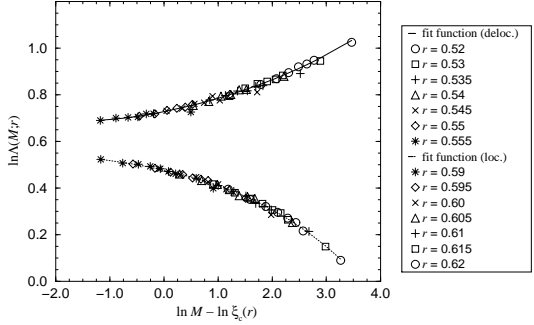}
  \end{center}
  \caption{Scaling function for the localization-delocalization
    transition. The upper branch corresponds to the delocalized and the
    lower to the localized phase.}
  \label{Skalenfunktion}
\end{figure}

Fig. \ref{Skalenfunktion} shows the scaling function with the upper
branch belonging to the metallic and the lower branch belonging to the
localized regime. The curves represent the fitted Chebyshev
polynomials. The data points are the raw data shifted by the fitted
values of $\xi_{\text{c}}(r)$. We omitted the data with $M=2$ in the
localized and $M=2$ and $M=4$ in the delocalized regime, because these
values showed systematic deviations due to finite-size effects. Also
data that are too close to the critical point were omitted. For the
remaining data the confidence test of the fit gives $\Delta_\Theta =
-0.30$ and $\Delta_\Theta = 0.27$ for the localized and delocalized
branch, respectively. So the assumption of one-parameter scaling is
very well confirmed.

Fixing the fit parameters $\ln\xi_{\text{c}}$ by setting
$\ln\xi_{\text{c}}(r=0.52)=0$ for the delocalized and
$\ln\xi_{\text{c}}(r=0.62)=0$ for the localized branch (circles in
Fig. \ref{Skalenfunktion}) the procedure has converged with an
accuracy of 0.1\% after about 50 iterations. The starting values have
a radius of convergence of about 5. So a rough estimate is sufficient
for convergence.

The Chebyshev polynomials used are of fourth order. With a lower order
it is impossible to fit the curves (as indicated by the figure of
merit $\Delta_\Theta$), whereas with an order higher than 6 the fitted
curves start to follow the fluctuations of the data points, which
results in a non-physical behavior of the scaling function. With an
order between four and six there is no significant difference in the
results.

\subsection{Critical Exponent $\bbox{\nu}$ and Critical RLL
  $\bbox{\Lambda}^\ast$} In order to determine the critical exponent
$\nu$ of the correlation length we used the fit procedure presented in
Sec. \ref{nu-Fit}. With a starting value $r^\ast\in[0.53,0.59]$ the
procedure converges. After about 10 iterations the corrections are
smaller than 1\%. The results of the fit are
\begin{equation}
  \nu = 2.51 \pm 0.18 \quad \text{and} \quad r^\ast = 0.571 \pm
  0.002\;.
\end{equation}
The prefactors take the values
\begin{mathletters}
  \begin{equation}
    \ln \xi_{\text{c}}^{\text{0,loc}} = -7.55 \pm 0.42
  \end{equation}
  and
  \begin{equation}
    \quad \ln \xi_{\text{c}}^{\text{0,deloc}} = -7.50- \pm 0.43.
  \end{equation}
\end{mathletters}
The confidence test of the fit gives $\Delta_\theta = -0.47$, showing
its high quality. It is very important to stress, that the given
errors (Eq. (\ref{E_theta})) are not independent. They have to be
interpreted considering the correlation matrix
\begin{equation}
  {\sf C}_{\theta} =
  \left ( \begin{array}{cccc}
    0.0310  & -0.0737  & -0.0739 & -0.0004\\
    -0.0737  & 0.1782  & 0.1766   & 0.0009\\
    -0.0739 & 0.1766    & 0.1853    & 0.0009\\
    -0.0004 & 0.0009 & 0.0009 & 5.5\cdot10^{-6}
  \end{array} \right ),
\end{equation}
which shows that the different values are highly correlated.

\begin{figure}
  \begin{center}
    \leavevmode
    \epsfxsize=7cm
    \epsffile{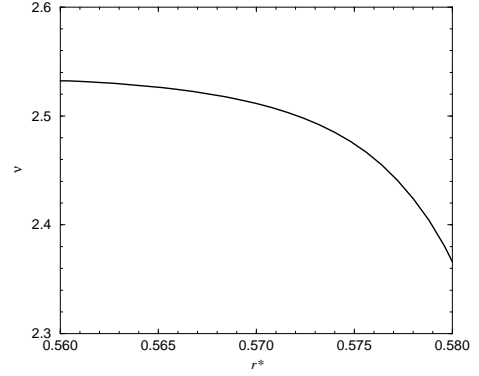}
  \end{center}
  \caption{Dependence of $\nu$ on the assumed value of $r^\ast$. The
    best estimate is found for $r^\ast=0.571$}
  \label{nu-rkrit}
\end{figure}

Several reference values for $\nu$ have been published in the last
years \cite{wegner89,zirn89,ando89,MacK_Local}. The most recent
calculations yield $\nu=2.75\pm0.1$ \cite{FABHRSWW91}, $\nu=2.5\pm0.3$
\cite{evang95} and $\nu=2.32\pm0.14$ \cite{SchZha97}. Within the
errors our value agrees with these. We note that $\nu$ is very
sensitive to slight variations of $r^\ast$. This is seen by fitting
$\nu$ for fixed values of $r^\ast$. As is shown in Fig.
\ref{nu-rkrit} $\nu$ changes by 30\% if $r^\ast$ changes by about 3\%.
The difficulties in obtaining a credible value for $\nu$ were the
reasons to employ the iterative fit procedure of Sec. \ref{nu-Fit}.

In order to determine the critical RLL we used Eqs.
(\ref{Lambda-non-log}) and (\ref{Lambda-krit}). The result is
\begin{equation}
  \Lambda^\ast = 1.83 \pm 0.03.
\end{equation}
Finally, Eq. (\ref{alpha_null}) yields the value for the scaling
exponent of the typical LDOS
\begin{equation}
  \alpha_0 = 2.174 \pm 0.003.
\end{equation}
This value agrees well with the result of Schweitzer \cite{schwe95},
$\alpha_0 = 2.19 \pm 0.03$ obtained for a Hamiltonian model.

\section{Summary}
\label{SUM}
In this work we found the S2NC-network model to be a new model to
describe mesoscopic disordered electron systems with symplectic
symmetry. We constructed the topology of the model and the scattering
matrices representing the potential and spin scatterers. Three
independent parameters were needed to characterize the scatterers. The
reflection coefficient $r$ and the transmission coefficient $t$
represent the strength of the spatial disorder by defining the mean
free path of the network model (in the absence of SOI). The
coefficient $s$ represents the strength of the spin-orbit scattering
and defines a corresponding spin scattering length. We have shown that
our model exhibits a localization-delocalization transition. In order
to investigate this transition we calculated renormalized localization
lengths $\Lambda(r,t,s)$ by the transfer matrix method and obtained
the scaling function by an iterative fit procedure. The quality of the
fit was checked by a $\chi^2$-test, which confirmed the assumption of
one-parameter scaling. We constructed a phase diagram for the system
showing a metallic phase for any $s>0$, if $r$ is small and $t$ is
large enough. The critical exponent of the correlation length was
obtained to be $\nu=2.51\pm0.18$. This value agrees with previously
published results within the errors. We pointed out, that in
determining the errors it is essential to take the correlations into
account. The critical renormalized localization length was found to be
$\Lambda^\ast = 1.83 \pm 0.03$.  By a conformal mapping argument this
corresponds to a value $\alpha_0 = 2.174 \pm 0.003$ for the scaling
exponent of the typical local density
of states.

\acknowledgments

We thank Peter Freche for valuable discussion and previous
collaboration. This work was performed within the research program of
the Sonderforschungsbereich 341 of the Deutsche
Forschungsgemeinschaft.


\end{document}